\documentclass[twocolumn, switch]{article} % Method A for two-column formatting

\usepackage{preprint}

\usepackage{amsmath, amsthm, amssymb}

\usepackage{xcolor}		% colors for hyperlinks
\usepackage[colorlinks = true,
            linkcolor = purple,
            urlcolor  = blue,
            citecolor = cyan,
            anchorcolor = black]{hyperref}	% Color links to references, figures, etc.
\usepackage{booktabs} 		% professional-quality tables
\usepackage{nicefrac}		% compact symbols for 1/2, etc.
\usepackage{microtype}		% microtypography
\usepackage{lineno}		% Line numbers
\usepackage{float}			% Allows for figures within multicol

\usepackage{lipsum}		%  Filler text

\usepackage{natbib}
\usepackage{algorithm}
\usepackage{algorithmicx}
\usepackage{algpseudocode}
\usepackage{graphicx}
\usepackage{tabularx}
 \usepackage{tikz}
\usepackage{placeins}
\usepackage{afterpage}
\usepackage{microtype}

\usepackage{newfloat}
\DeclareFloatingEnvironment[name={Supplementary Figure}]{suppfigure}
\usepackage{sidecap}
\sidecaptionvpos{figure}{c}

\usepackage{titlesec}
\titlespacing\section{0pt}{12pt plus 3pt minus 3pt}{1pt plus 1pt minus 1pt}
\titlespacing\subsection{0pt}{10pt plus 3pt minus 3pt}{1pt plus 1pt minus 1pt}
\titlespacing\subsubsection{0pt}{8pt plus 3pt minus 3pt}{1pt plus 1pt minus 1pt}

\usepackage{tikz,xcolor,hyperref}

\definecolor{lime}{HTML}{A6CE39}
\DeclareRobustCommand{\orcidicon}{
	\begin{tikzpicture}
	\draw[lime, fill=lime] (0,0) 
	circle [radius=0.16] 
	node[white] {{\fontfamily{qag}\selectfont \tiny ID}};
	\draw[white, fill=white] (-0.0625,0.095) 
	circle [radius=0.007];
	\end{tikzpicture}
	\hspace{-2mm}
}
\foreach \x in {A, ..., Z}{\expandafter\xdef\csname orcid\x\endcsname{\noexpand\href{https://orcid.org/\csname orcidauthor\x\endcsname}
			{\noexpand\orcidicon}}
}
\usepackage[english]{babel}
\title{Generation of patient specific cardiac chamber models using generative neural networks under a Bayesian framework for electroanatomical mapping}

\usepackage{watermark}
\thiswatermark{
\put(170,-705){\color{gray}{*correspondence: \texttt{sunil.mathew@marquette.edu}}}
}
\usepackage{authblk}

\author[1\thanks{\tt{sunil.mathew@marquette.edu}}]{Sunil Mathew\orcidA{}}
\author[2]{Jasbir Sra\orcidB{}}
\author[1]{Daniel B. Rowe\orcidC{}}

\affil[1]{Department of Mathematical and Statistical Sciences, Marquette University, Milwaukee, WI-53233}
\affil[2]{Electrophysiology, Aurora Healthcare, Milwaukee, WI-53215}

\begin{document}

\twocolumn[ % Method A for two-column formatting
\begin{@twocolumnfalse} % Method A for two-column formatting
  
\maketitle

\begin{abstract}
  Electroanatomical mapping is a technique used in cardiology to create a detailed 3D map of the electrical activity in the heart. It is useful for diagnosis, treatment planning and real time guidance in cardiac ablation procedures to treat arrhythmias like atrial fibrillation. A probabilistic machine learning model trained on a library of CT/MRI scans of the heart can be used during electroanatomical mapping to generate a patient-specific 3D model of the chamber being mapped. The use of probabilistic machine learning models under a Bayesian framework provides a way to quantify uncertainty in results and provide a natural framework of interpretability of the model. Here we introduce a Bayesian approach to surface reconstruction of cardiac chamber models from a sparse 3D point cloud data acquired during electroanatomical mapping. We show how probabilistic graphical models trained on segmented CT/MRI data can be used to generate cardiac chamber models from few acquired locations thereby reducing procedure time and x-ray exposure. We show how they provide insight into what the neural network learns from the segmented CT/MRI images used to train the network, which provides explainability to the resulting cardiac chamber models generated by the model. 
\end{abstract}
%\keywords{First keyword \and Second keyword \and More} % (optional)
\vspace{0.35cm}

\end{@twocolumnfalse} % Method A for two-column formatting
] % Method A for two-column formatting

%\begin{multicols}{2} % Method B for two-column formatting (doesn't play well with line numbers), comment out if using method A

%%%%%%%%%%%%%%%  Main text   %%%%%%%%%%%%%%%
% \linenumbers
%% main text
\section{Introduction}
Electroanatomic mapping (EAM) is part of Electrophysiology studies (EPS) which are minimally invasive procedures that are useful for diagnosis, treatment planning and real time guidance in cardiac ablation procedures to treat arrhythmias like atrial fibrillation which affects millions of people in the United States. It is estimated that over 12 million people in the United States will be affected by atrial fibrillation by 2030 \cite{colilla2013estimates}. It is characterized by rapid and irregular beating of the atria (upper chambers of the heart) which can lead to blood clots, stroke, heart failure and other heart-related complications. Cardiac ablation is a procedure used to treat atrial fibrillation by eliminating the heart tissue that causes the abnormal heart rhythm. EAM is performed by an electrophysiologist prior to ablation to create a 3D map of the heart's electrical activity to identify the location of the abnormal heart tissue. The electrophysiologist uses this map to guide the catheter to the location of the abnormal heart tissue and eliminate it using radiofrequency (RF) energy or cryogenic energy.

During an RF ablation procedure, the electrode at the tip of the catheter is placed in contact with the tissue causing the arrhythmia, and high-frequency electrical energy is delivered to create a localized lesion or scar. The goal of RF ablation is to disrupt the abnormal electrical pathways in the heart causing the arrhythmia and create a new pathway that follows the normal electrical conduction system of the heart, thus restoring normal heart rhythm and reducing or eliminating the need for medication to control the arrhythmia. 

Minimizing the x-ray exposure from the fluoroscope to the patient is one of the electrophysiologist's goals while performing electroanatomical mapping. Using a probabilistic machine learning model in an electroanatomical mapping system can provide a good approximation of the mapped chamber with a few acquired locations from the mapping catheter in the chamber of interest. This considerably reduces the time taken to map the chamber, thus minimizing the x-ray exposure from the fluoroscope.

\section{Electroanatomical mapping systems}

An EAM system is a state-of-the-art medical device that is used for navigation during cardiac ablation procedures and to create a detailed and accurate map of the electrical activity in a patient's heart. The system typically consists of a computer, high-definition monitors that are strategically positioned in front of the electrophysiologist during the procedure, a specialized catheter with electrodes in its distal end that captures electrical information, and specialized software that is used to analyze signals and create a 3D electroanatomical map of the chamber of interest.

EAM systems vary based on the technology that is used to capture the location of the catheter. The commonly used technologies are magnetic location tracking \cite{lior1997}, impedance based location tracking and image processing \cite{sra2016identifying}. EAM systems that are based on magnetic location tracking places a device under the patient table to generate a magnetic flux which is used to locate the catheter. EAM systems that use impedance based tracking utilizes body surface patches to measure the impedance and locate the catheter. EAMs that use image processing techniques tracks the catheter in fluoroscope images to converts them to 3D location. 

The high accuracy and precision of EAM systems have revolutionized the diagnosis and treatment of various cardiac arrhythmias. However the maps produced using modern electranatomical systems do not produce anatomically accurate chamber models when only few points are acquired. The shape produced is representative of the points acquired and does not use any prior information pertaining to the shape of the chamber being mapped and may not produce an accurate anatomical model \cite{Mukherjee2014}. They only start resembling the chamber of interest when a large number of points are acquired covering all areas of the chamber. This is a major drawback of the current electroanatomical mapping systems. We propose a surface reconstruction method that uses probabilistic machine learning models to generate a detailed anatomical model of the chamber of interest from a few points acquired by the mapping catheter. This will help in reducing the time taken to map the chamber and thus reduce the x-ray exposure from the fluoroscope.

\section{Method}

\begin{figure}[!hb]
  \centering
  \includegraphics[width=\linewidth]{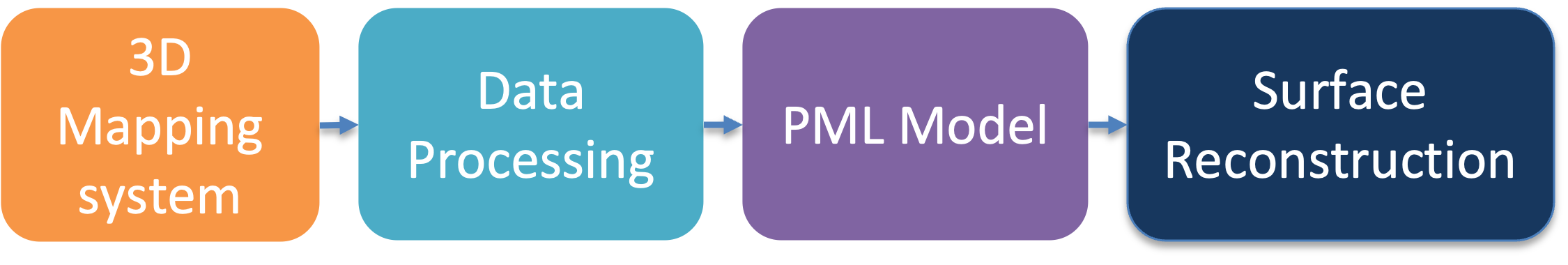}
  \caption{Generation of 3D cardiac chamber models}
  \label{fig:3dsrs}
  \end{figure}

There are four main steps involved in our approach to generate a 3D cardiac chamber model from 3D location and electrical information acquired by an electroanatomical mapping system. The block diagram shown in figure \ref{fig:3dsrs} visualizes the information flow in the system. The data collected by the 3D mapping system is passed on to the Data processing step, then to the trained Probabilistic Machine Learning Model, and the output of the machine learning model is fed into the surface reconstruction module in the last step to generate a patient specific cardiac model. Each block is explained in detail in following sections.

\subsection{Data processing}

The points acquired during electroanatomical mapping are in the coordinate space of the mapping system. The 3D location information acquired during mapping is in a raw unstructured format. The data is preprocessed to transform it into voxel data \cite{Xu2021} that can be fed into the machine learning model. Voxels are the 3D equivalent of pixels in 2D images. They are the smallest unit of a 3D image. The 3D location data is transformed into voxel space using the following equation

\begin{equation*}
    \label{eq:voxel}
    v = \frac{p - p_{min}}{p_{max} - p_{min}} * n
\end{equation*}

where $p$ is the 3D location of the point, $p_{min}$ and $p_{max}$ are the minimum and maximum 3D location values defined by the field-of-view (FOV), $n$ is the number of voxels in each dimension and $v$ is the voxel location. The number of voxels in each dimension is chosen based on the number of voxels in each dimension of the training data used to train the machine learning model. Mapping systems tend to match the FOV of the fluoroscope to project 3D models onto the 2D fluoroscope image.

The next step is compute the convex hull of all acquired points. A Delaunay 3D triangulation algorithm with an alpha value which produces a closed surface can be used to get an alpha shape \cite{Edelsbrunner94}. The alpha value is the radius of a ball that determines whether a simplex (vertex, edge, face or tetrahedron) in the Delaunay triangulation is part of the alpha shape or not. An alpha value of zero results in a convex hull. The resulting simplices of the Delaunay triangulation can be used to estimate which voxels are inside versus outside the alpha shape. The voxels inside are set to 1 and the voxels outside are set to zero. Different values of alpha can be chosen to obtain a concave hull that might produce different results, but they were not explored as they may cause the surface to have holes with higher values of alpha. The resulting voxels are then passed onto the machine learning model to generate inference.

\subsection{Probabilistic Machine Learning Model}

Probabilistic Machine Learning models are models that use probability theory and statistics to represent uncertainty and variability in data and the learning processes. They can be used for tasks such as classification, regression, clustering, anomaly detection, generative modeling and reinforcement learning. Unlike regular neural networks they are more interpretable. Here they are used in the context of generative modeling. \\

Unlike a discriminative model which attempts to learn a boundary between different kinds of data points, generative models aim to learn the underlying structure and distribution of the data, and then generate new data that resembles the original data. They can capture complex and high-dimensional patterns in the data that are difficult to model explicitly. The goal here is to learn a representation of the chamber of interest (left atrium of the heart). We will be using latent variable models for this purpose. At the core of generative modeling is Bayesian inference, which uses Baye's rule to obtain the posterior distribution of the parameters of the data model,

\begin{equation*}
  \label{eq:bayes}
P(\theta|D)=\frac{P(D|\theta).P(\theta)}{P(D)}
\end{equation*}

where $\ P(D|\theta) $ is the data likelihood with parameters $\theta$ of the model, $P(\theta)$ is the prior probability, and $\ P(D)$ is the evidence which can be computed as sum over all possible values of $ \theta $. This distribution is then utilized to generate inference that can provide insight into the learning process and the model. When there is a large number of parameters in the model which is the case in neural networks, the denominator term in \eqref{eq:bayes} becomes intractable.\\

This marginal likelihood is calculated using the sum and product rule of probability. This denominator term known as evidence is a normalizing constant which can be omitted to get the relationship,
\begin{equation*}
P(\theta|\ D)\propto\ P(D|\theta)P(\theta).
\end{equation*}

These formulas above form the core of Bayesian inference. The posterior distribution is proportional to the product of the prior and the likelihood. The prior distribution is the distribution of the parameters before seeing the data. The likelihood is the probability of the data given the parameters. The posterior distribution is the distribution of the parameters after seeing the data.\\

In a generative model where latent variables $z$ are a subset of parameters $\theta$. The following equation represents the generative model, where $\tilde{x}$ is the generated data, $\mathbf{f}$ is a function that maps a latent variable $\mathbf{z}$ to the generated data.

\begin{equation*}
\tilde{x} = \mathbf{f}(\mathbf{z})
\end{equation*}

The prior distribution of the latent variable $\mathbf{z}$ is represented by the equation below. 

{%
\abovedisplayskip=0pt
\begin{align*}
\mathbf{z} &\sim p(\mathbf{z}) 
\end{align*}
\belowdisplayskip=0pt
}%

The prior distribution is used to sample from the latent space to generate new data. A new data point $\mathbf{x}$ can be used to update the prior distribution of the latent variable $\mathbf{z}$ using Bayes' rule. The posterior distribution is given by,

\begin{equation*}
  p(\mathbf{z} \mid \mathbf{x}) = \frac{p(\mathbf{x} \mid \mathbf{z}) p(\mathbf{z})}{p(\mathbf{x})}
\end{equation*}

where $p(\mathbf{x} \mid \mathbf{z})$ is the likelihood. This posterior distribution is intractable due the high dimensional integral posed by the large number of parameters in the neural network. To address this problem we employ approximate Bayesian inference techniques. The two main techniques of approximate Bayesian inference are Markov chain Monte Carlo (MCMC) methods and Variational Bayes, also known as Variational Inference (VI). In the following sections we describe how segmented CT/MRI images can be modeled using approximate Bayesian inference techniques and how the results compare in the context of generating patient specific cardiac models. First we present the data and then the two different approximate Bayesian inference techniques that can be employed to generate patient specific cardiac models.

\subsubsection{Training Data}

In probabilistic machine learning, training data is essential for building and fine-tuning models. The quality and quantity of training data directly impact the performance of the model, including its accuracy, precision, and generalizability. These models use training data to learn patterns and relationships between the input features and the output variable.

Training data that is biased, incomplete, or inaccurate can negatively impact the model's performance. Therefore, it's essential to carefully select and preprocess the training data to ensure its quality and relevance to the problem at hand. Left atrial anatomy can be highly variable. Its important that every data point in the dataset has the same amount of pulmonary vein segmented along with the left atrial cavity. The datasets also needs to be normalized as some scans may have a different field of view or voxel size. Voxels belonging to both the left atrial cavity and the myocardium need to be segmented together as one class. 

\begin{figure}[!h]
  \centering
  \includegraphics[scale=.8]{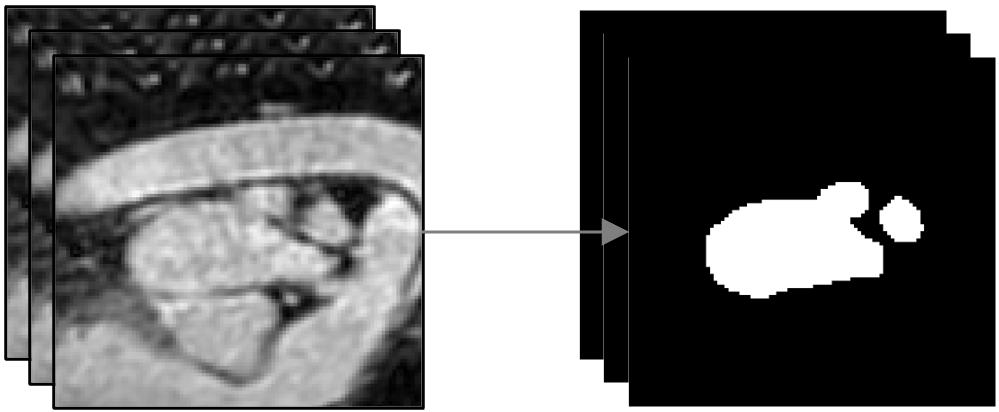}
  \caption{Segmented Left Atrial data}
  \label{fig:training_data_la}
\end{figure}

The training data used to train the PML model consists of binary voxels with zeros for the background and ones for the left atrium and pulmonary veins. The individual slices stack up to become a $n_x \times \ n_y \times \ n_z$ voxel grid to form one 3D data point.

\subsubsection{Restricted Boltzmann Machine Model for Segmented Left Atrial Data}

Restricted Boltzmann Machines (RBM) \cite{Smolensky1986, Freund92, CD2002} are energy based stochastic artificial neural networks that can learn a probability distribution over its inputs. The structure of an RBM for segmented left atrial data is shown below in Figure \ref{fig:rbm}. 
\begin{figure}[!h]
  \centering
  \includegraphics[scale=.8]{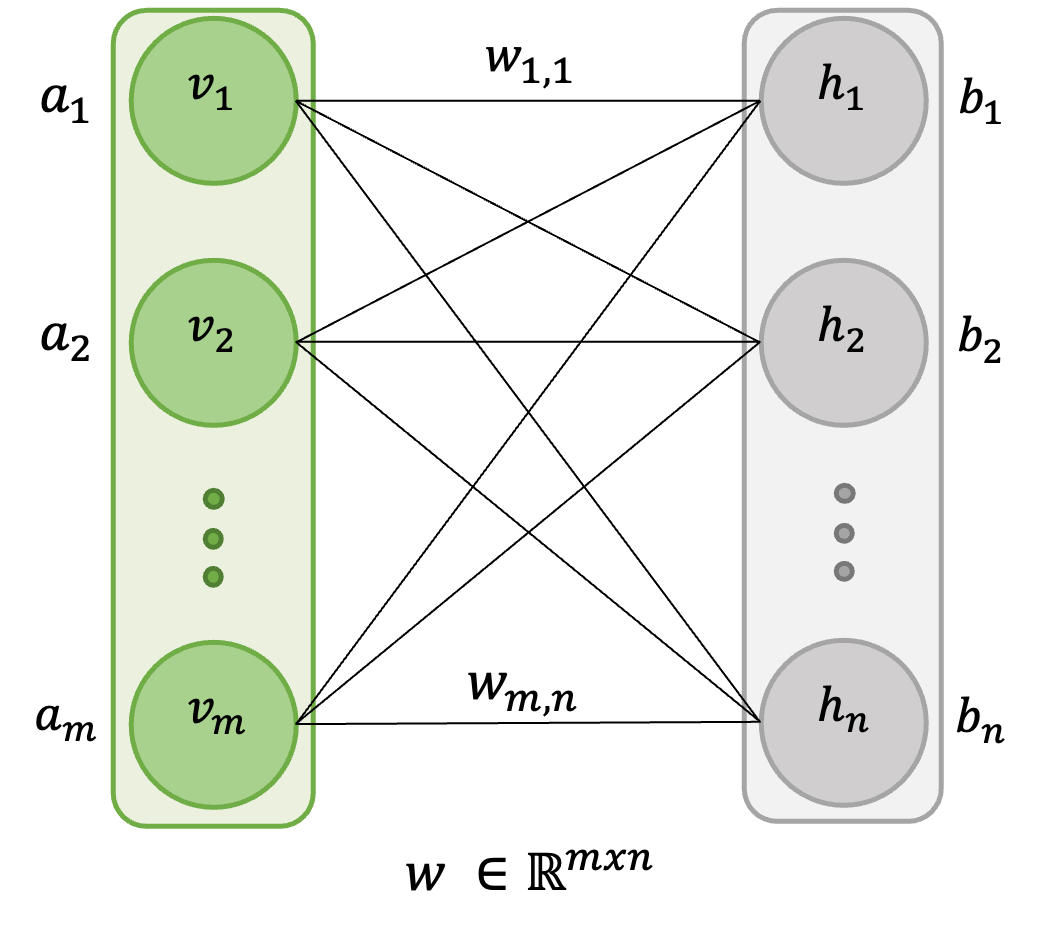}
  \caption{Structure of a Restricted Boltzmann machine}
  \label{fig:rbm}
\end{figure}
It consist of binary visible ($v$) and hidden ($h$) neurons connected by a $m \times n$ weight matrix $W_{m \times n}$. There are no interconnections between the visible nodes or the hidden nodes and they form a bipartite graph. There is a vector of bias $a_m$ associated with the visible nodes and another vector of bias $b_m$ associated with the hidden nodes. The energy of a pair of $v$, $h$ is given by the following,
\begin{equation*}
E(v,\ h)=-\sum_{i=1}^{m}{a_iv_i-\sum_{j=1}^{n}{b_jh_j}}-\ \sum_{i=1}^{m}\sum_{j=1}^{n}{w_{ij}v_ih_j}
\end{equation*}
\begin{equation*}
E(v,\ h)=-a^Tv-b^Th-\ v^TWh.
\end{equation*}

Here the visible nodes represent the voxels of a segmented CT/MRI data and the hidden nodes represent features learned from the dataset. There can be three kinds of parameters introduced into this network. The parameter $w_{ij}$ for strength between a visible and hidden node, $b_i$ for every visible node, $c_j$ for every hidden node. The joint probability distribution of this MRF specifies the probability the network assigns to a pair of visible and hidden vector and can be written as,
\begin{equation*}
P(v,\ h)=\frac{1}{Z}e^{-E(v,\ h)}
\end{equation*}
where the partition function sums over all such pairs of $v$, $h$,
\begin{equation*}
Z=\sum_{V}\sum_{H} e^{-E(v,\ \ h)}
\end{equation*}
and is the normalizing constant to make the probability distribution sum to 1. The probability for an image data or visible vector can be obtained by summing over all the possible hidden vector configurations 
\begin{equation*}
P(v)=\frac{1}{Z}\sum_{H} e^{-E(v,\ h)} .
\end{equation*}
When there are more than a few hidden units, it becomes difficult to compute the partition function as it will have exponentially many terms. So Markov chain Monte Carlo methods such as Gibbs sampling is used to obtain samples from the model starting from a global configuration. The conditional distributions for the hidden and visible units are used to construct a Markov chain and run until it reaches its stationary distribution. The probability of a global configuration at thermal equilibrium is an exponential function of its energy. Since there are no connection between nodes in a group they are independent. The conditional probability for visible nodes given hidden nodes are
\begin{equation*}
P(v| h)=\prod_{i}^{m}{P(v_i|h)}.
\end{equation*}
This probability values allow us to know how much of the current configuration of hidden nodes are represented by the image and conversely the conditional probability for hidden nodes given the visible nodes are 
\begin{equation*}
P(h| v)=\prod_{j}^{n}{P(h_j|v)}.
\end{equation*}
From the above probabilities we can also estimate the probability of one of the visible or hidden node being activated. Since there is no connection between visible nodes,

\begin{equation*}
P(h_j=1| v)=\sigma(\sum_{i=1}^{m}{w_{ij}v_i+b_j\ })
\end{equation*}
\begin{equation*}
P(h| v)=\prod_{j=1}^{n}{\sigma{{(v}^TW_{:j}+b}_j)}=\sigma{{(v}^TW_{:i}+b}_j)
\end{equation*}
where $\sigma$ denotes the logistic sigmoid. Similarly, 
\begin{equation*}
P(v_i=1| h)=\sigma(\sum_{j=1}^{n}{w_{ij}h_j+a_i\ })
\end{equation*}
\begin{equation*}
P(v| h)=\prod_{i=1}^{m}{\sigma{{(v}^TW_{:i}+a}_i)}.
\end{equation*}

\subsubsection{Training the RBM Model}

The goal of training a RBM to be a generative model is to maximize the likelihood of the training dataset of images. This may prove to be difficult but \cite{CD2002} showed that there is a less obvious objective function than log likelihood of the data to optimize called the Contrastive Divergence which is the difference between two KL divergences. It consists of performing block Gibbs sampling and batch updates to the weights and biases of the network similar to stochastic gradient descent in a regular neural network that uses back propagation for training. The weights and biases of the network are adjusted to lower the energy of the training image and raise the energy of other images which are low in energy, thereby making a large contribution to the partition function.

{%
\abovedisplayskip=0pt
\begin{figure}[!ht]
  \centering
  \includegraphics[scale=.5]{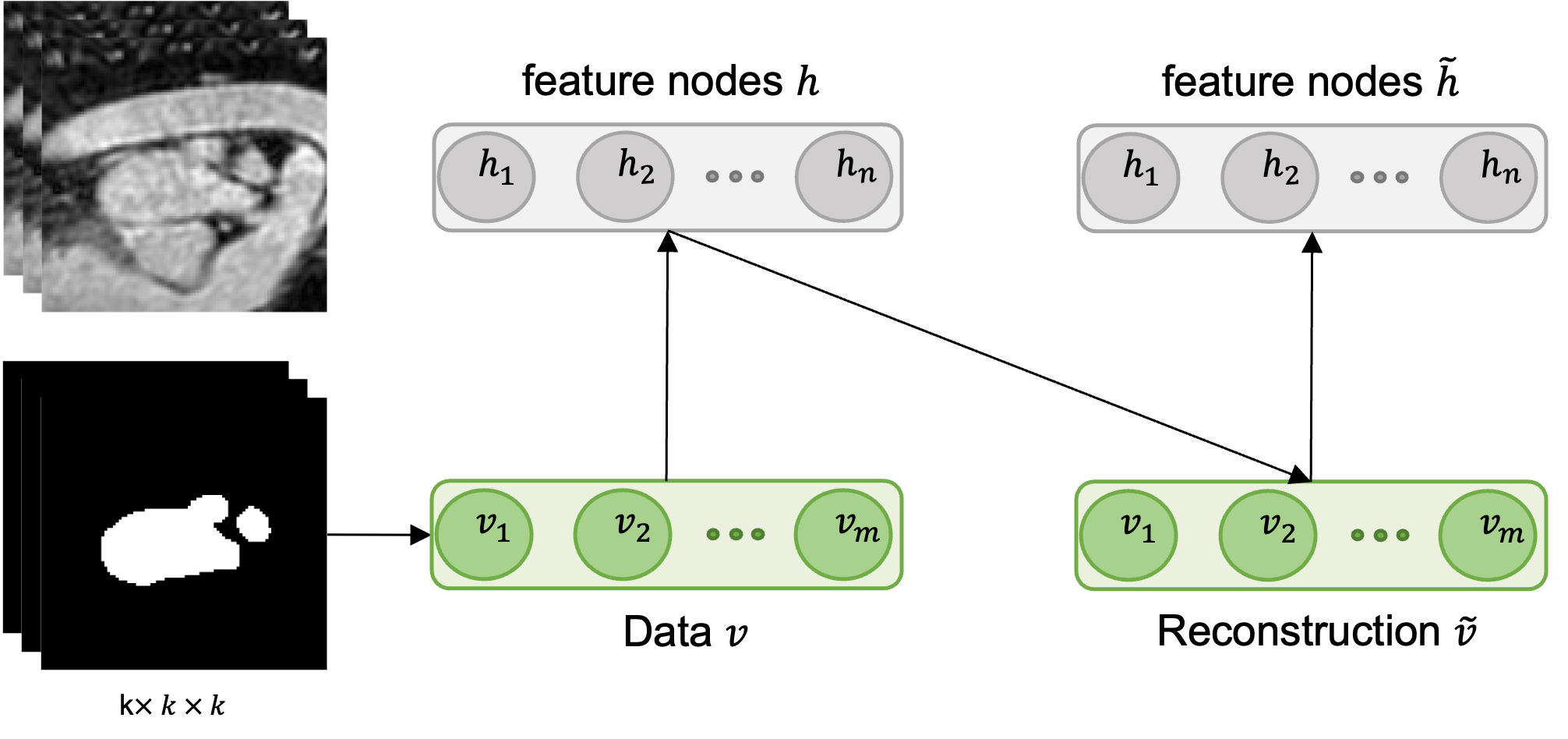}
  \caption{Single training step of the model}
  \label{fig:training}
\end{figure}
\belowdisplayskip=0pt
}%

The gradient of the log-likelihood with respect to the weights is given by

{%
\abovedisplayskip=0pt
\begin{equation*}
  \frac{\partial}{\partial w_{ij}} \log P(v) = \langle h_i v_j \rangle_{\text{data}} - \langle h_i v_j \rangle_{\text{recon}}
\end{equation*}
\belowdisplayskip=0pt
}%

where $\langle.\rangle_{\text{data}}$ and $\langle.\rangle_{\text{recon}}$ denote the expectation over the data and the model distributions, respectively. The second term in this equation is difficult to compute. 

\begin{algorithm}[H]
\caption{k-Contrastive Divergence algorithm}
\label{alg:kcd}

\begin{algorithmic}[1]
\Require $\mathbf{v}$: visible units, $\mathbf{W}$: weights, $\mathbf{b}$: visible bias, $\mathbf{c}$: hidden bias, $k$: number of Gibbs sampling steps, $\alpha$: learning rate
\While{not converged}
\For{$i = 1$ to $n_{batches}$} \Comment{For each batch (v) of images}
\State $\mathbf{h} \gets \sigma(\mathbf{W} \cdot \mathbf{v}^{(i)} + \mathbf{c})$ \Comment{Positive phase}
\State $\mathbf{v}_0 \gets \mathbf{v}^{(i)}$
\For{$j = 1$ to $k$} \Comment{Negative phase}
\State $\mathbf{v}_j \gets \sigma(\mathbf{W}^T \cdot \mathbf{h} + \mathbf{b})$
\State $\mathbf{h} \gets \sigma(\mathbf{W} \cdot \mathbf{v}_j + \mathbf{c})$
\EndFor
\State $\mathbf{W} \gets \mathbf{W} + \alpha \cdot (\mathbf{v}_0 \cdot \mathbf{h}^T - \mathbf{v}_k \cdot \mathbf{h}_k^T)$ 
\State $\mathbf{b} \gets \mathbf{b} + \alpha \cdot (\mathbf{v}_0 - \mathbf{v}_k)$
\State $\mathbf{c} \gets \mathbf{c} + \alpha \cdot (\mathbf{h}^{(0)} - \mathbf{h}_k)$
\EndFor
\EndWhile
\end{algorithmic}

\end{algorithm}

In contrastive divergence learning instead of starting with a randomly initialized visible vector and running the Gibbs chain for a large number of steps, a data point from the training set is used as the initial value for the visible vector and $k$ steps of Gibbs sampling is used to get a reconstruction ($k=1$ works well) to get the second term from what is known as the negative phase of contrastive divergence learning. The difference between the positive and negative phase is used to update the weights and biases. The algorithm is summarized in Algorithm \ref{alg:kcd}. The update rule seen in step (9) in the algorithm above can be written as,

\begin{equation*}
\Delta w_{ij} = \epsilon \left(\langle v_i h_j \rangle_{data} - \langle v_i h_j \rangle_{recon}\right)
\end{equation*}

where $\epsilon$ is the learning rate, $w_{ij}$ is the weight connecting the visible unit $i$ and hidden unit $j$, $\langle v_i h_j \rangle_{data}$ is the expected value of the product of the visible unit $i$ and hidden unit $j$ under the data distribution, and $\langle v_i h_j \rangle_{recon}$ is the expected value of the product of the visible unit $i$ and hidden unit $j$ under the model distribution. The update rule for the biases are given by,

\begin{align*}
\Delta b_i &= \epsilon \left(\langle v_i \rangle_{data} - \langle v_i \rangle_{recon}\right) \\
\Delta c_j &= \epsilon \left(\langle h_j \rangle_{data} - \langle h_j \rangle_{recon}\right)
\end{align*}

where $\langle v_i \rangle_{data}$ is the expected value of the visible unit $i$ under the data distribution, and $\langle v_i \rangle_{recon}$ is the expected value of the visible unit $i$ under the model distribution. The algorithm is run until the weights and biases converge to produce accurate reconstructions of the segmented data. The weights and biases represent what the model has learned from the training data and is utilized for inference.

\subsubsection{Model Interpretability and Inference}

\begin{figure}[!h]
  \centering
  \includegraphics[width=\linewidth]{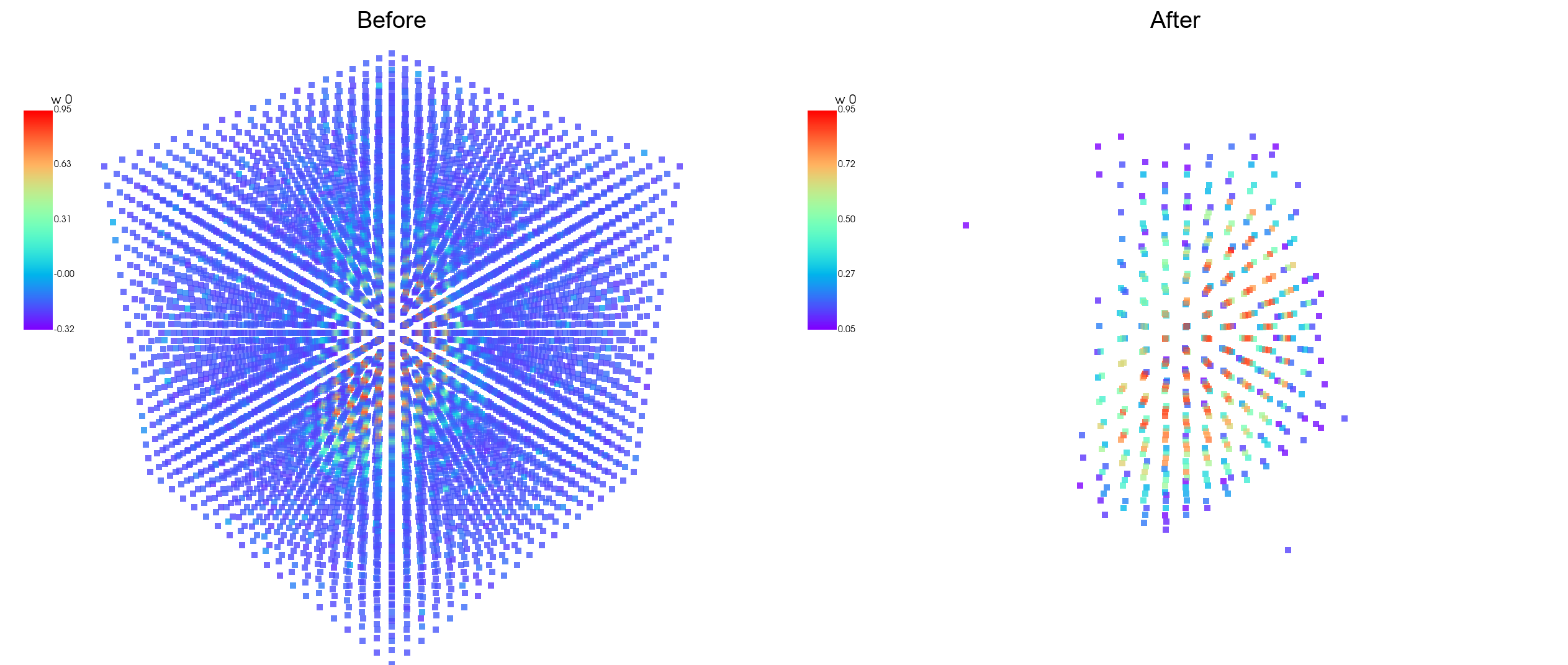}
  \caption{3D RBM weights before(left) and after(right) pruning.}
  \label{fig:ml_trained_weights_pruning}
\end{figure}

After the training procedure, it is possible to examine each feature and infer what the model has learned from the training dataset. As each voxel is connected to all nodes in the feature vector, so we can represent the weights as a 3D voxel grid. Figure \ref{fig:ml_trained_weights_pruning} shows weights of one of the hidden nodes represented as a voxel grid, before and after pruning low magnitude weights. The voxels in the cavity of the left atrium are of higher magnitude suggesting that the model has more certainty on classifying them as being part of the foreground or the left atrium. 

Different nodes within the network learn distinct features, and there may be correlations among these features. When there are only few hidden nodes, each node has a more significant impact on the representation of the volume data. The hidden nodes are forced to work together and correlate their activity to reconstruct the volume data effectively. Consequently, the features learned by the RBM tend to have higher interdependence and can exhibit strong correlations. However, as the number of hidden nodes increases, each hidden node has more freedom to capture different aspects of the data independently. With more nodes, the hidden layer can distribute the workload among them, allowing different nodes to specialize in capturing different aspects of the data. This can lead to a reduction in the correlation between features and improve quality of reconstruction.

After training the RBM, we can obtain the posterior predictive distribution by sampling from the posterior distribution of the hidden nodes.

The posterior probabilities of hidden nodes given visible nodes, trained weights, hidden biases are calculated as
\begin{equation*}
p(h_j = 1|v, W, c) = \sigma\left(\sum_i W_{ij}v_i + c_j\right)
\end{equation*}
where $h_j$ is the $j$-th hidden unit.

A binary sample from the posterior distribution of hidden nodes are drawn as
\begin{equation*}
h_{s_{ij}} \sim \mathrm{Bernoulli}\left(p(h_j = 1|v, W, c)\right)
\end{equation*}
where $h_{s_{ij}}$ is the $i$-th sample of the $j$-th hidden unit.

A sample of a hidden vector is drawn as

\begin{equation*}
  h_s \sim \mathrm{Bernoulli}(p(h|v, W, c))
\end{equation*}

The posterior probabilities of visible nodes given hidden nodes are calculated as
\begin{equation*}
p(v_i = 1|h_s, W, b) = \sigma\left(\sum_j W_{ij}h_{s_{ij}} + b_i\right).
\end{equation*}

The mean and standard deviation of the posterior predictive distribution are calculated as
\begin{equation*}
v_{mean} = \frac{1}{n}\sum_{s=1}^{n} p(v|h_s, W, b)
\end{equation*}

\begin{equation*}
  v_{std} = \frac{1}{n} \sqrt{\sum_{s=1}^{n} \left(p(v|h_s, W, b) - \mathrm{v_{mean}}\right)^2}.
\end{equation*}

where $n$ is the number of samples drawn from the posterior distribution of the hidden nodes. The mean and standard deviation of the posterior predictive distribution are used to reconstruct the volume and quantify the uncertainty in the reconstruction.

\begin{figure}[!h]
  \centering
  \includegraphics[width=\linewidth]{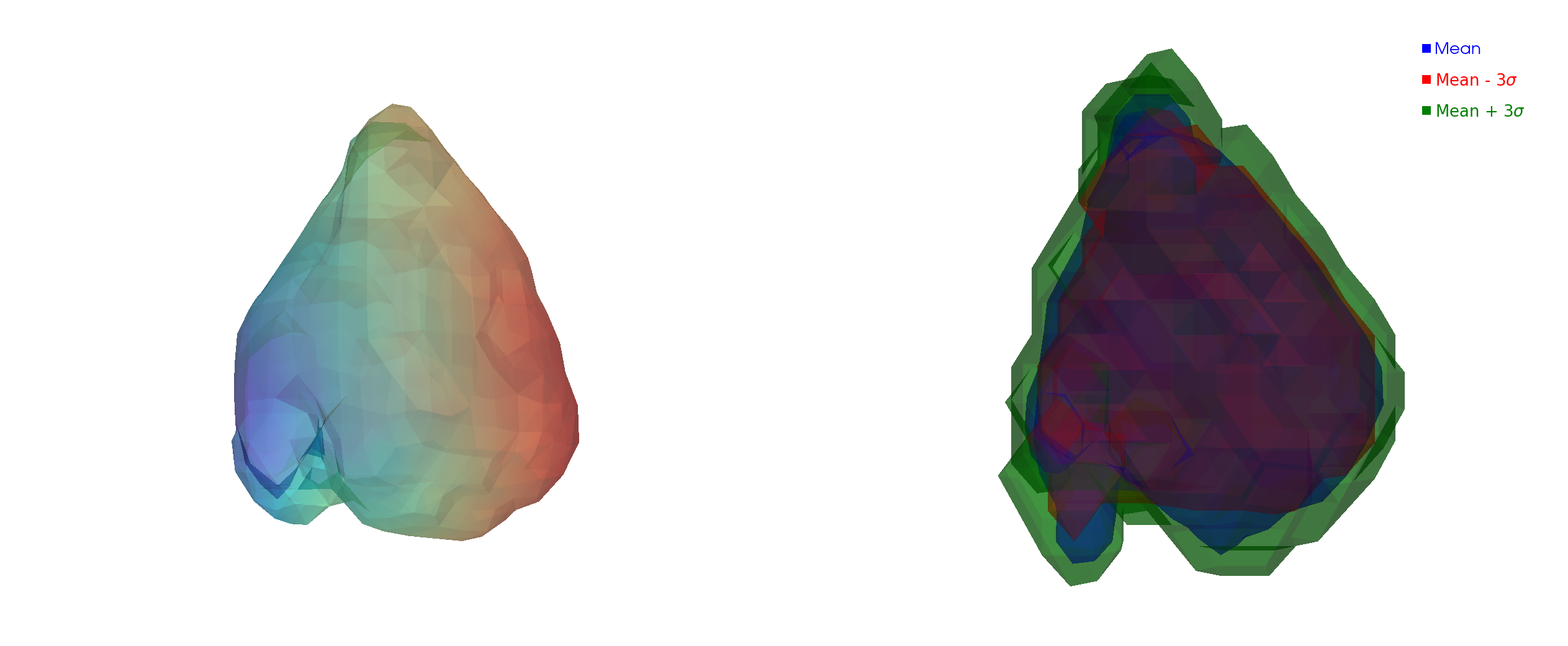}
  \caption{True surface (left)mean(blue), mean+std (green), mean-std (red) surfaces from RBM model output}
  \label{fig:rbm_aleatoric}
\end{figure}

Figure \ref{fig:rbm_aleatoric} shows the mean, mean +/- standard deviation surfaces reconstructed using posterior probability distribution of the hidden nodes of an RBM.

\subsubsection{A Variational Autoencoder (VAE) model for volumetric data}

MCMC methods allow for the computation of posterior distributions with high accuracy, as they asymptotically converge to the true posterior distribution. They are able to sample from complex, high dimensional distributions with multiple modes and non-convex shapes. However, they come with significant computational costs, that is where Variational Inference (VI) techniques can prove to be useful. Variational Inference \cite{peterson1987a,jordan1999an} aims at posing the intractable inference problem as one of optimization. It approximates the posterior distribution of a latent variable model by an easy to sample distribution such as a Gaussian distribution. 

A Variational Autoencoder (VAE) \cite{kingma2014autoencoding,rezende2014stochastic} is a generative model that learns a compressed latent representation of the data by mapping it to a latent space defined by a prior probability distribution. It consists of an encoder network that maps the data to latent variables, and a decoder network that reconstructs the data from the latent variables. The main difference between a VAE and a standard autoencoder \cite{hinton2006reducing} is that the latent space in a VAE is stochastic. The output of the encoder is a distribution over the latent vectors, rather than a single vector. This allows the VAE to generate new data points by sampling from the latent space.

\begin{figure}[h!]
  \includegraphics[width=\linewidth]{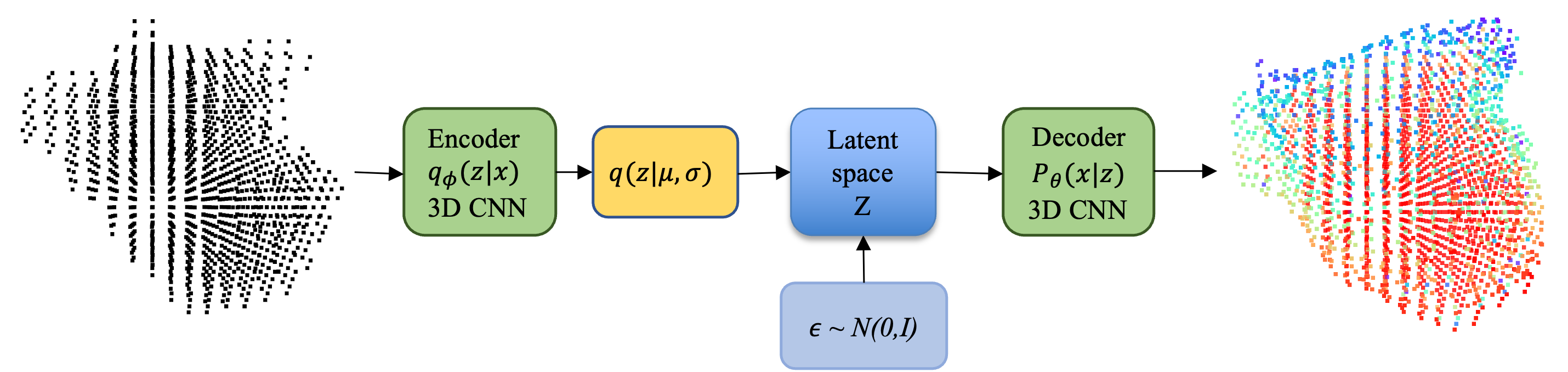}
  \caption{3D Variational Autoencoder Model}
  \label{fig:vae_3d}
\end{figure}

VAEs are trained to minimize the difference between the input and the reconstructed output, and the divergence between the learned latent variable and a known prior distribution. Several methods and architectures have been developed to use a VAE for 3D data. One of the most popular methods is the 3D Convolutional VAE (3D-CVAE) \cite{Wu2016} used for modeling volumetric objects. The 3D-CVAE is similar to a traditional VAE, but uses 3D convolutions instead of 2D convolutions to process 3D data. They are good at capturing local spatial information in the image data. Figure \ref{fig:vae_3d} shows the block diagram of a VAE that is used to reconstruct segmented CT or MRI data.

Here we try to learn a compressed representation of segmented CT or MRI data $x$ that is optimized to match a multivariate Gaussian distribution with mean $\mu$ and variance $\sigma$. This is achieved by introducing a latent variable $z$ that represents the compressed representation of the volume (eg: left atrium), and training the VAE to learn a conditional probability distribution $p_{\theta}(x|z)$ that maps the compressed representation $z$ to the input data $x$. The parameters $\theta$ of the conditional probability distribution are learned during training.

The goal is to learn a compressed representation $z$ that maximizes the joint probability $p_{\theta}(x|z)p(z)$ of the data $x$ given the compressed representation $z$ and the prior distribution $p(z)$ over the compressed representation. Maximizing this joint probability ensures that the learned representations are both good at reconstructing the original data ($p_{\theta}(x|z)$ is high) and align well with our prior belief about what $z$ should be like ($p(z)$ is high).

However, this joint distribution is intractable, as it involves integrating over all possible values of $z$. To overcome this problem, the VAE introduces an approximate posterior distribution $q_{\phi}(z|x)$ which can be used to approximate the true posterior distribution $p(z|x)$.

The VAE learns the parameters $\phi$ of the approximate posterior distribution $q_{\phi}(z|x)$ by minimizing the KL divergence between the approximate posterior and the true posterior. The KL divergence is defined as:

\begin{equation*}
KL(q_{\phi}(z|x) || p(z|x)) = \int q_{\phi}(z|x) \log\frac{q_{\phi}(z|x)}{p(z|x)} dz
\end{equation*}

Minimizing the KL divergence ensures that the approximate posterior distribution is as close as possible to the true posterior distribution. The loss function for the VAE is the sum of two terms: the reconstruction loss and the KL divergence loss. The reconstruction loss measures the difference between the input data and the data reconstructed from the compressed representation $z$ using the decoder. It is defined as:

\begin{equation*}
\mathcal{L}{\text{rec}}(\theta,\phi;x) = -\mathbb{E}{q_{\phi}(z|x)}[\log p_{\theta}(x|z)].
\end{equation*}

The KL divergence loss measures the difference between the approximate posterior distribution and the prior distribution:

\begin{equation*}
\mathcal{L}{\text{KL}}(\phi;x) = KL(q{\phi}(z|x) || p(z)).
\end{equation*}

The total loss function is the sum of the reconstruction loss and the KL divergence loss:

\begin{equation*}
\mathcal{L}(\theta,\phi;x) = \mathcal{L}{\text{rec}}(\theta,\phi;x) + \mathcal{L}{\text{KL}}(\phi;x).
\end{equation*}

which is equivalent to maximizing the ELBO:

\begin{equation*}
\mathcal{L}(\theta,\phi;x) = \mathbb{E}_{q_{\phi}(z|x)}[\log p_{\theta}(x|z)] - KL(q_{\phi}(z|x) || p(z)).
\end{equation*}

During training, the parameters $\theta$ and $\phi$ are learned by minimizing the total loss function with respect to these parameters:

\begin{equation*}
\theta, \phi = \underset{\theta, \phi}{\arg\min} \mathcal{L}(\theta, \phi;x).
\end{equation*}

Training via backpropagation is made possible in VAEs using the reparameterization trick. It involves reparameterizing the latent variables $z$ as a function of a random noise variable $\epsilon$ that can be sampled from a standard normal distribution. Specifically, we can express $z$ as

\begin{equation*}
z = \mu + \sigma \odot \epsilon
\end{equation*}

where $\mu$ and $\sigma$ are the mean and standard deviation of the approximate posterior distribution $q_{\phi}(z|x)$, and $\odot$ denotes element-wise multiplication.

By reparameterizing $z$ in this way, we can sample from the approximate posterior distribution $q_{\phi}(z|x)$ using a simple and differentiable transformation. This allows us to backpropagate through the sampling process, which is necessary for training the VAE using stochastic gradient descent. 

This results in a model that can learn meaningful representations of the left atrium from segmented CT or MRI data, which then can be used to generate patient-specific models of the left atrium by sampling from the learned latent space when presented with seed points acquired during mapping.

We can generate a new data point $x$ by first sampling a latent variable $z$ from the prior distribution $p(z)$, and then passing it through the decoder to get a generated output $\tilde{x}$

\begin{equation*}
z \sim p(z), \quad \tilde{x} = p_\theta(x|z).
\end{equation*}

Samples from $p_\theta(x|z)$ can be used to obtain a mean surface as well as a measure of uncertainty. The mean surface is obtained by taking the mean of the samples, and the uncertainty is obtained by taking the standard deviation of the samples

\begin{equation*}
\tilde{x} = \mu_{z \sim p(z)}(p_\theta(x|z)), \quad \sigma_{z \sim p(z)}(p_\theta(x|z)).
\end{equation*}

\subsubsection*{Visualizing the Latent Space in 3D}

\begin{figure}[!h]
  \includegraphics[width=\linewidth]{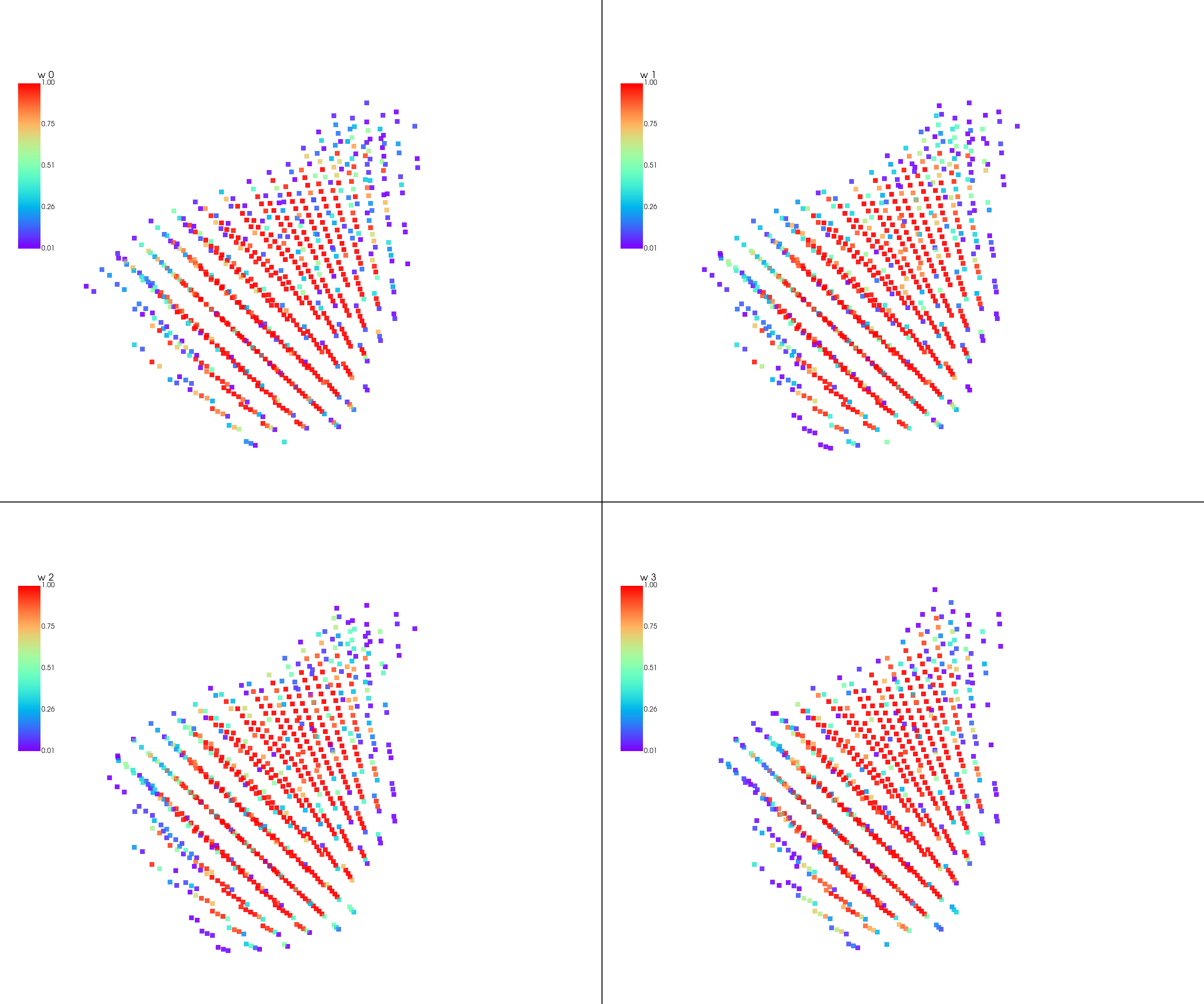}
  \caption{3D Latent Space of a VAE trained on segmented left atrial data.}
  \label{fig:vae_latent_space}
\end{figure}

The Figure \ref{fig:vae_latent_space} shows few representations from the 3D latent space of a VAE trained on segmented MRI images of the left atrium of the heart. The latent space of a VAE is typically high-dimensional, when a latent space is 2D or 3D, it can be visualized by sampling from a uniform 2D or 3D grid. Similarly when the latent space has a dimensionality greater than three we can use the following method to visualize the latent space from a n-dimensional uniform grid.

A standard normal distribution is used to sample k equally spaced elements from each dimension of the latent space defined by the n-dimensional grid. 

\begin{equation*}
  d_i = q(linspace(a, b, k))
\end{equation*}

where $q(\cdot)$ is the quantile function of the standard normal distribution, $a$ and $b$ are the lower and upper bounds of each dimension of the latent space, $k$ is the number of linearly spaced points obtained by the following equation,

\begin{equation*}
\text{linspace}(a, b, k)  = \{a, a + \frac{(b - a)}{k-1}, a + 2\frac{(b - a)}{k-1}, \ldots, b\}
\end{equation*}

Suppose we have an n-dimensional latent space, $ D_1, D_2, \dots, D_n$, then a Cartesian product of the n elements from each dimension can be taken to obtain all the possible ordered combinations of the n elements from each dimension. 

% \begin{equation*}
%   \text{Cartesian Product}(D_1, D_2, \ldots, D_n) = (d_1, d_2 \ldots, d_n) \mid d_1 \in D_1, \ldots, d_n \in D_n
% \end{equation*}

\begin{equation*}
  \begin{aligned}
  \text{Cartesian Product}(D_1, D_2, \ldots, D_n) &= (d_1, d_2 \ldots, d_n) \mid \\
  & d_1 \in D_1, \ldots, d_n \in D_n
  \end{aligned}
\end{equation*}

From these ordered combinations, a linearly spaced set of points are sampled to obtain a set of points in the latent space which is then be passed through the decoder to obtain a visualization of the latent space. The latent space provides insights into the distribution of the data and the learning process. It can be used to identify outliers and anomalies in results generated by the model. This can guide the practitioner to either update the training data or update the model itself.

\subsection*{Surface Reconstruction}

The next step in our system is Surface Reconstruction. The dense point cloud generated by the ML model is used to generate a 3D surface mesh model of the chamber of interest. The surface reconstruction algorithm used in our system is the 3D Marching Cubes algorithm \cite{lorensen1987marching}, which is a popular method for generating a 3D surface mesh from volumetric data. 

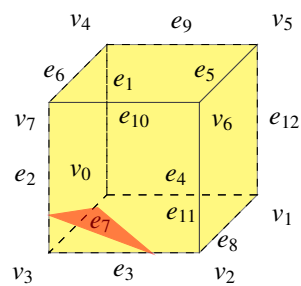
\begin{figure}[htbp]
  \centering
\begin{tikzpicture}
  % define the dimensions of the cube
  \pgfmathsetmacro{\cubex}{2}
  \pgfmathsetmacro{\cubey}{2}
  \pgfmathsetmacro{\cubez}{2}
  % draw the visible faces of the cube
  \draw[black,fill=yellow,opacity=0.5] (0,0,0) coordinate (v000) -- ++ (-\cubex,0,0) coordinate (v100) -- ++ (0,-\cubey,0) coordinate (v110) -- ++ (\cubex,0,0) coordinate (v010) -- cycle;
  \draw[black,fill=yellow,opacity=0.5] (0,0,0) -- ++ (0,0,-\cubez) coordinate (v001) -- ++ (0,-\cubey,0) coordinate (v011) -- ++ (0,0,\cubez) -- cycle;
  \draw[black,fill=yellow,opacity=0.5] (0,0,0) -- ++ (-\cubex,0,0) -- ++ (0,0,-\cubez) coordinate (v101) -- ++ (\cubex,0,0) coordinate (v001) -- cycle;
  % draw the hidden faces of the cube
  \draw[black,dashed] (-\cubex,-\cubey,-\cubez) coordinate (v111) -- ++(\cubex,0,0) -- ++(0,\cubey,0) -- ++(-\cubex,0,0) -- cycle;
  \draw[black,dashed] (-\cubex,-\cubey,-\cubez) -- ++(0,\cubey,0) -- ++(0,0,\cubez) -- ++(0,-\cubey,0) -- cycle;
  \draw[black,dashed] (-\cubex,-\cubey,-\cubez) -- ++(\cubex,0,0) -- ++(0,0,\cubez)  -- ++(-\cubex,0,0) -- cycle;
  % label the vertices of the cube
  \node [below right=1pt] at (v000) {$v_{6}$};
  \node [below right=2pt] at (v010) {$v_{2}$};
  \node [below left=2pt] at (v110) {$v_{3}$};
  \node [below left=1pt] at (v100) {$v_{7}$};
  \node [above right=2pt] at (v001) {$v_{5}$};
  \node [below right=2pt] at (v011) {$v_{1}$};
  \node [above left=2pt] at (v111) {$v_{0}$};
  \node [above left=2pt] at (v101) {$v_{4}$};
  % label the edges of the cube
  \path [midway] (v000) -- node[above=1pt] {$e_1$} (v100);
  \path [midway] (v100) -- node[left=1pt] {$e_2$} (v110);
  \path [midway] (v110) -- node[below=1pt] {$e_3$} (v010);
  \path [midway] (v010) -- node[left=1pt] {$e_4$} (v000);
  \path [midway] (v000) -- node[left=1pt] {$e_5$} (v001);
  \path [midway] (v100) -- node[left=1pt] {$e_6$} (v101);
  \path [midway] (v110) -- node[right=1pt] {$e_7$} (v111);
  \path [midway] (v010) -- node[below=1pt] {$e_8$} (v011);
  \path [midway] (v001) -- node[above=1pt] {$e_9$} (v101);
  \path [midway] (v101) -- node[right=1pt] {$e_{10}$} (v111);
  \path [midway] (v111) -- node[below=1pt] {$e_{11}$} (v011);
  \path [midway] (v011) -- node[right=1pt] {$e_{12}$} (v001);

  % define the vertices of the triangle
  \coordinate (A) at (-2, -1.5, 0);
  \coordinate (B) at (-1.35, -1.4, 0);
  \coordinate (C) at (-0.5, -1.9, 0.3);
  % draw the triangle inside the cube
  \draw[red,fill=red,opacity=0.5] (A) -- (B) -- (C) -- cycle;
  % label the vertices of the triangle
  % \node [right=2pt] at (A) {$A$};
  % \node [above left=2pt] at (B) {$B$};
  % \node [right=2pt] at (C) {$C$};
  % end of tikzpicture
  \end{tikzpicture}
  \caption{A voxel with an isosurface facet.}
  \label{fig:voxel}
\end{figure}

The dense 3D point cloud voxel data is divided into small cubes, each of which is then examined to determine whether it contains any parts of the left atrial surface. The algorithm considers each of the eight vertices of the cube to determine whether it lies inside or outside the surface. The scalar value at each vertex (voxel value) is compared to a threshold value. If the scalar value is greater than the threshold, the vertex is considered to be inside the surface, otherwise it is marked to be outside. The algorithm then uses the results to determine which edges of the cube are intersected by the surface and connects them to form a triangle as seen in Figure \ref{fig:voxel}. The algorithm then repeats this process for each cube in the volume.

\begin{figure}[!h]
\centering
\includegraphics[width=\linewidth]{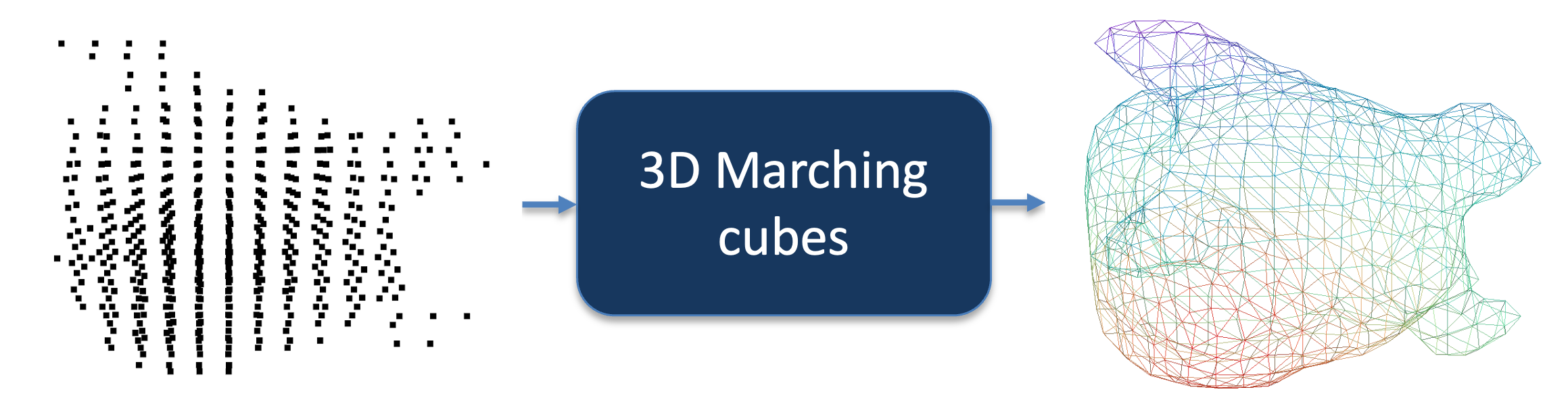}
\caption{3D Surface mesh generated using marching cubes on ML output.}
\label{fig:3d_mc}
\end{figure}

The resulting triangles are then connected to form a surface mesh. Figure \ref{fig:3d_mc} shows a 3D surface mesh generated using marching cubes on the output of the machine learning model. Here the output is the mean from the posterior distribution of the latent variable model. Other summary statistics such as variance can be used to visualize the uncertainty in the model prediction. This can be useful for guiding the electrophysiologist to regions of the chambers where more points need to acquired to get a better approximation of the chamber.

\subsection*{Post processing}

Post processing is an important step in 3D mesh generation pipeline, as it can significantly improve the quality and usability of the mesh. Here the 3D mesh generated from the 3D marching cubes algorithm is first passed through a filter which removes small, isolated mesh components. Then it is passed through a smoothing filter and finally checked for holes and filled if any is found to get a watertight mesh that represents the left atrium.

\subsection*{Evaluation Metrics}

The evaluation metrics used to compare the accuracy of reconstructions of the data is the dice score. The dice score measures the similarity between the ground truth and the generated data. The dice score is calculated as follows:

\begin{equation*}
Dice = \frac{2 \times |A \cap B|}{|A| + |B|}
\end{equation*}

where $A$ and $B$ are the ground truth and the generated data, respectively. The dice score is a value between 0 and 1, where 1 is the best possible score.

\section{Left Atrial Dataset}

Left atrial anatomy can be highly variable across patients. While size and shape can vary, the main feature that can be used to categorize them is the number of pulmonary veins (PV). The most common type of left atrium has 2 left (LPV) and 2 right (RPV), other variants are 2 LPVs and 3 RPVs, and 1 LPV and 2 RPVs. \cite{krum2013left}.

The data used in the experiments in the following sections are from Task02\_Heart dataset \cite{Task02} of the Medical Segmentation Decathlon held as part of the Medical Image Computing and Computer Aided Interventions (MICCAI) Conference 2018 at Granada, Spain. This dataset was originally released through the Left Atrial Segmentation Challenge (LASC) 2013. The dataset has 20 segmented MRI scans with a voxel resolution of $1.25 \times 1.25 \times 2.7 mm^3$ covering the entire heart acquired during a single cardiac phase (free breathing with respiratory and ECG gating). The MRI scans were segmented by experts to obtain ground truth segmentation labels. All of the datasets have four pulmonary veins (~74\% of the population). Out of the total of 20 scans, 15 were employed for training purposes, while the remaining 5 scans were utilized for testing.

\subsection{Experiments}

For all experiments, the MSD dataset was cropped and downsampled to be a $20 \times 20 \times 20$ 3D array of voxels. The system was trained for 100 epochs with a batch size of 1 for both the RBM and VAE based systems to generate inference. Point acquisition by the electroanatomical mapping system was simulated by randomly sampling points from the test data. The simulation involved generating the true surface using marching cubes, then sampling $n$ vertices from the resulting surface mesh. The vertices are then compared to voxel locations that were used to generate the mesh by measuring their Euclidean distance. The closest voxel location is then recorded as a point acquired during mapping. Algorithm \ref{alg:eam_simulation} summarizes the steps involved in simulating electroanatomical mapping. 

\begin{algorithm}[h!]
  \caption{EAM Simulation}
  \label{alg:eam_simulation}
  \begin{algorithmic}[1]
    \Require 3D data point (voxel data) from the test set
    \Require Number of points to be acquired ($n$)
    \Ensure Points acquired during mapping

    \State \textbf{Generate true surface:}
    \State Generate true surface $s_{true}$ using marching cubes

    \State \textbf{Sample points from true surface $s_{true}$:}
    \State $v_n \leftarrow$ sample $n$ random points from the vertices of $s_{true}$
    \For{each point $p$ in the vertex list $v_n$}
      \For{each point $p_v$ in the voxel data}
        \State Calculate Euclidean distance $d$ between $p$ and $p_v$
        \If {$d <= \text{threshold}$}
          \State Add $p_v$ to points list
        \EndIf
      \EndFor
    \EndFor

    \State \textbf{Return the points list}
  \end{algorithmic}
\end{algorithm}

Three EP studies were simulated, the first where $n=25$ points were acquired, the second where $n=100$ points were acquired, and the third where $n=100$ points were acquired. The points were then passed on to the system to generate a 3D surface mesh model of the left atrium. The 3D surface mesh model was then compared to the ground truth data using dice scores to evaluate the accuracy of the system.

\section{Results}

The results from the three simulated experiments are discussed in this section. Table \ref{tab:rbm_vae_dice_score} shows the dice scores generated by the two systems for 20, 100, 250 points,

\begin{figure*}[!h]
  \centering
  \includegraphics[width=\linewidth]{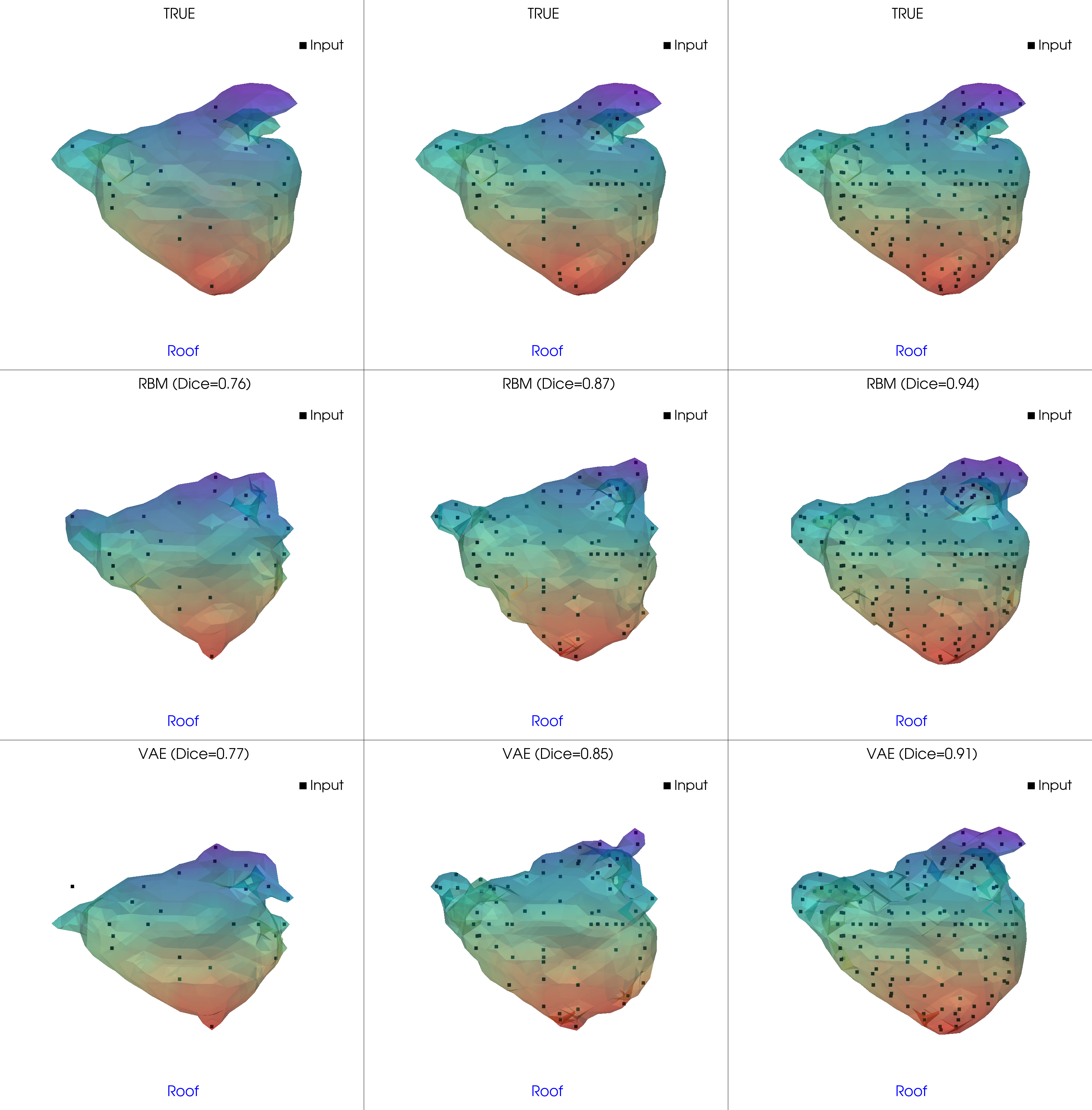}
  \caption{RBM, VAE comparison plot with 25 (left), 100 (middle) and 250 (right) points in Roof view.}
  \label{fig:rbm_vae_comparsion_roof}
\end{figure*}

\begin{figure*}[!h]
  \centering
  \includegraphics[width=\linewidth]{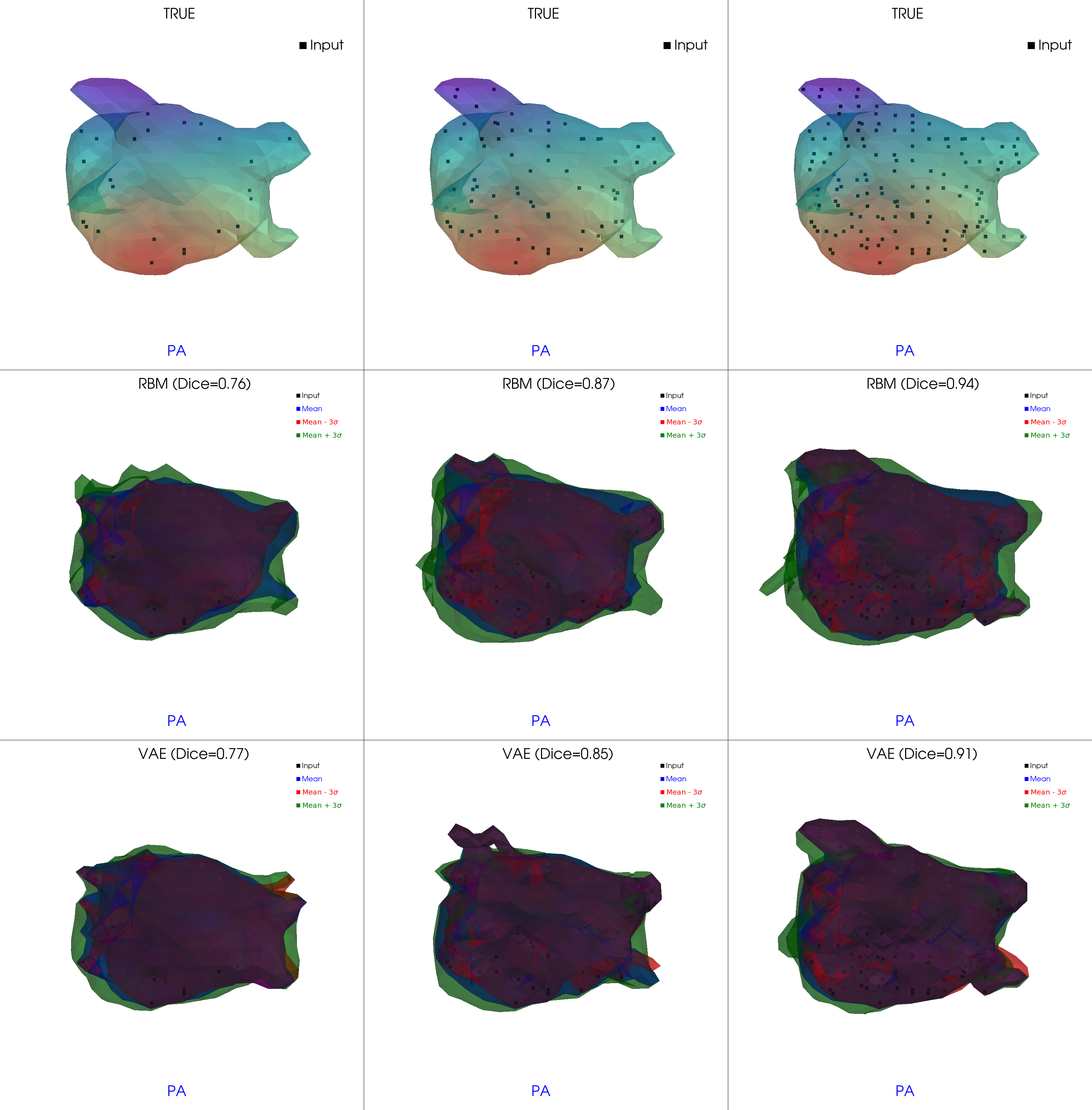}
  \caption{RBM, VAE uncertainty comparison with 20, 50 and 100 points in PA view.}
  \label{fig:rbm_vae_alea_comparsion_pa}
\end{figure*}

\begin{table}[h!]
\centering
\caption{Comparison of dice scores for RBM and VAE based systems.}
\begin{tabular}{|c|c|c|c|}
\hline
\textbf{Model} & \textbf{20 points} & \textbf{100 points} & \textbf{250 points} \\ \hline
RBM & 0.79 & 0.89 & 0.92 \\ \hline
VAE & 0.76 & 0.83 & 0.88 \\ \hline
\end{tabular}
\label{tab:rbm_vae_dice_score}
\end{table}

From Table \ref{tab:rbm_vae_dice_score}, we can observe that the dice score increases as the number of points acquired increases. We can also observe that the RBM based system performs better than the VAE based system. This is because the RBM based system is able to generate a more accurate approximation of the ground truth data than the VAE based system.

Figure \ref{fig:rbm_vae_comparsion_roof} shows comparison of surface reconstruction results using the RBM, VAE systems for a patient data from the test set in Roof view. The first row shows the surface reconstructed using the ground truth data and the points that were sampled from it for the reconstructions in the second and third row. The first column shows reconstructions using 25 points, second column shows reconstructions using 100 points, and the third column shows reconstructions using 250 points. The RBM based system is able to generate a more accurate approximation of the chamber compared to the VAE system even though they don't differ much in dice scores. As expected and seen from the dice scores in table \ref{tab:rbm_vae_dice_score}, the reconstructions are better with more points acquired.

The Figure \ref{fig:rbm_vae_alea_comparsion_pa} shows the uncertainty of the RBM, VAE model outputs in the Posterior-Anterior (PA) view. 100 samples were generated to reconstruct the mean, mean +/- standard deviation surfaces. It can be observed that both RBM, VAE models exhibit more uncertainty in the areas where fewer points were acquired. There is also high uncertainty in areas where the pulmonary veins are located in the model. This is due to the high variability of pulmonary veins across patients in the training set.

\begin{table}[h!]
  \centering
  \caption{Task-based assessment of five patient datasets by three expert observers.}
 \resizebox{\linewidth}{!}{ \begin{tabular}{|p{1.2cm}|p{0.5cm}|p{0.5cm}|p{0.5cm}|p{0.5cm}|p{0.5cm}|p{0.5cm}|p{0.5cm}|p{0.5cm}|p{0.5cm}|p{0.5cm}|p{0.5cm}|p{0.5cm}|p{0.5cm}|p{0.5cm}|p{0.5cm}|p{0.5cm}|p{0.5cm}|p{0.5cm}|p{0.5cm}||p{0.5cm}|p{0.5cm}|} 
    \hline Patient & \multicolumn{6}{c|}{RBM} & \multicolumn{6}{c|}{VAE}\\ 
    \cline{2-13} data & \multicolumn{3}{c|}{Visual quality} & \multicolumn{3}{c|}{Match w true} & \multicolumn{3}{c|}{Visual quality} & \multicolumn{3}{c|}{Match w true} \\ 
    \cline{2-13} & 25 & 100 & 250 & 25 & 100 & 250 & 25 & 100 & 250 & 25 & 100 & 250 \\ 
  \hline 1 & 3.0 & 3.3 & 4.0 & 2.3 & 3.6 & 4.6 & 2.3 & 3.3 & 3.3 & 2.0 & 3.6 & 5.0 \\
  \hline 2 & 3.6 & 3.6 & 4.0 & 3.0 & 3.3 & 4.6 & 2.6 & 3.3 & 4.3 & 3.3 & 3.0 & 4.0 \\
  \hline 3 & 3.0 & 3.3 & 4.6 & 2.0 & 3.3 & 4.0 & 2.3 & 3.0 & 3.3 & 2.3 & 3.6 & 4.3 \\
  \hline 4 & 3.0 & 3.0 & 5.0 & 2.6 & 3.6 & 3.6 & 2.0 & 4.0 & 3.6 & 3.0 & 3.3 & 3.6 \\
  \hline 5 & 2.3 & 3.6 & 4.3 & 2.0 & 4.0 & 4.3 & 1.6 & 3.0 & 4.0 & 2.3 & 3.0 & 4.0 \\
  \hline Average & 3.0 & 3.4 & 4.4 & 2.4 & 3.6 & 4.2 & 2.2 & 3.3 & 3.7 & 2.6 & 3.3 & 4.2 \\
  \hline 
  \end{tabular} }
  \label{tab:task_score} 
\end{table}

Table \ref{tab:task_score} shows task-based assessment of the surface reconstructions of the five patient datasets by three expert observers. Observers individually scored the visual quality and conformity with the true anatomy of the generated meshes on a scale of 1 to 5, where 1 is the worst possible score and 5 is the best possible score. The table shows the average score of the three observers for the two systems. The table shows that both systems are capable of producing useful reconstructions with the RBM based system performing slightly better than the VAE based system. The table also shows that the reconstructions improve with the number of points acquired. 

\section{Conclusion}

Generating anatomical models of the left atrium from a sparse point cloud of acquired locations is a challenging task, but one that is essential for Electrophysiology studies. A probabilistic machine learning approach to the problem provides a way to bring uncertainty quantification and interpretability to a neural network solution to the problem. A Bayesian approach provides greater insight into the model which incorporates prior knowledge from CT/MRI data. Such a model can be useful for navigation during electroanatomical mapping. 

The results of the conducted experiments show that the proposed system is capable of generating anatomically accurate models of the left atrium from a sparse point cloud of few acquired locations. This can enable creating quick maps during electrophysiology studies and reduce the time taken for mapping during a study, thereby reducing the patient's exposure to radiation from the fluoroscope.

%%Harvard
\bibliographystyle{model2-names}
\bibliography{refss}

\end{document}